\documentclass[%
reprint,  
 amsmath,amssymb,
 aps,
]{revtex4-2}
\usepackage{enumitem}
\usepackage{graphicx}
\usepackage{dcolumn}
\usepackage{bm}
\usepackage{hyperref}
\usepackage{slashed}
\usepackage{subfloat}
\usepackage{multirow}
\usepackage[table]{xcolor}
\usepackage{pgfplots,subfigure}
\usepackage{orcidlink}
\usepackage{float}
\usepackage{geometry}
\usepackage[table]{xcolor} 
\usepackage{booktabs}       
\usepackage{array}          
\usepackage{hyperref}
\usepgfplotslibrary{fillbetween}
\usepackage{colortbl,xcolor} 
\usepackage{tikz}
\usetikzlibrary{arrows.meta}

\newcolumntype{C}{>{\centering\arraybackslash}m{4cm}} 
\definecolor{acsblue}{RGB}{17,76,139}

\begin{document}

\fontsize{7.6}{8.6}\selectfont
\title{Electromagnetic wave propagation in static black hole spacetimes: an effective refractive index description in Schwarzschild geometry }

\author{Abdullah Guvendi\orcidlink{0000-0003-0564-9899}}
\email{abdullah.guvendi@erzurum.edu.tr }
\affiliation{Department of Basic Sciences, Erzurum Technical University, 25050, Erzurum, Türkiye}

\author{Omar Mustafa\orcidlink{0000-0001-6664-3859}}
\email{omar.mustafa@emu.edu.tr }
\affiliation{Department of Physics, Eastern Mediterranean University, 99628, G. Magusa, north Cyprus, Mersin 10 - Türkiye}

\author{Semra Gurtas Dogan\orcidlink{0000-0001-7345-3287}}
\email{semragurtasdogan@hakkari.edu.tr (Corr. Author)}
\affiliation{Department of Medical Imaging Techniques, Hakkari University, 30000, Hakkari, Türkiye}

\author{Hassan Hassanabadi\orcidlink{0000-0001-7487-6898}}
\email{hha1349@gmail.com}
\affiliation{Department of Physics, University of Hradec Kr$\acute{a}$lov$\acute{e}$, Rokitansk$\acute{e}$ho 62, 500 03 Hradec Kr$\acute{a}$lov$\acute{e}$, Czechia}
\affiliation{Department of Physics and Electronics, Khazar University, 41 Mahsati Str, 1096 Baku, Azerbaijan}

\date{\today}

\begin{abstract}
{\fontsize{7.6}{8.6}\selectfont \setlength{\parindent}{0pt} \textcolor{black}{We investigate electromagnetic wave propagation in static, spherically symmetric black hole spacetimes using a covariant and gauge-invariant framework based on the established Maxwell perturbation formalism. Building upon known parity decompositions and gauge-invariant master equations, we reformulate the resulting radial dynamics entirely within Schwarzschild coordinates and introduce an effective refractive-index description of electromagnetic propagation in curved spacetime.} Starting from the source-free Maxwell equations on a curved background, electromagnetic perturbations are decomposed according to parity and systematically reduced to gauge-invariant dynamical variables without introducing auxiliary coordinate transformations or horizon-regular variables. Both axial and polar sectors are shown to obey the same parity-independent master equation, and their exact isospectrality emerges naturally as a direct consequence of Maxwell theory in four dimensions. By eliminating first-derivative terms through an appropriate field redefinition, the radial dynamics is cast into a Helmholtz-type equation, which motivates the introduction of an effective, position- and frequency-dependent refractive index encoding gravitational redshift, curvature effects, and angular momentum within a unified optical framework. Specializing to the Schwarzschild geometry, we obtain the refractive index in closed analytical form and analyze its behavior in the near-horizon, intermediate, and asymptotic regimes. The resulting description provides a transparent and physically intuitive interpretation of electromagnetic evanescence, and propagation in black hole spacetimes, and establishes a robust foundation for wave-optical, semiclassical, and numerical studies in more general static gravitational backgrounds.}
\end{abstract}

\keywords{Electromagnetic perturbations, Schwarzschild spacetime, Gauge invariance, Wave propagation, Optical analogy, Refractive index}

\maketitle


\section{Introduction}
\label{sec:intro}

\setlength{\parindent}{0pt}

The propagation of electromagnetic fields in curved spacetime occupies a fundamental position in both classical and quantum aspects of gravitational physics. As a linear massless gauge field with a well defined covariant formulation, electromagnetism provides a uniquely controlled setting in which the influence of spacetime geometry on wave dynamics can be examined with full mathematical and physical precision \cite{thorne1982electrodynamics,mashhoon1973scattering,hollowood2009refractive,moncrief1974gravitational,regge1957stability}. The study of black hole perturbations across a range of spin fields, including spin-1 electromagnetic fields, has revealed rich dynamical phenomena such as scattering, absorption, greybody factors, and quasinormal ringing, which encode detailed information about spacetime geometry and boundary conditions at the event horizon and spatial infinity \cite{BertiCardosoStarinets_2009_review,CardosoLemos_PRD64_2001,Mamani2022_EPJC}. Early studies of the classical scattering and absorption of electromagnetic waves by Schwarzschild black holes established foundational results for wave cross sections and phase shifts in curved backgrounds \cite{Fabbri1975}. The quasinormal mode spectrum of electromagnetic perturbations, which governs the characteristic damped oscillations following perturbations of black holes, has been extensively analyzed in asymptotically flat, anti-de Sitter, and modified gravity contexts \cite{CardosoLemos_PRD64_2001,CardosoLemos_PRD64_2001,Konoplya2005}. In addition to quasinormal spectra, greybody factors that quantify deviations from pure black body emission due to backscattering by the gravitational potential have been studied in both four-dimensional and higher-dimensional black hole backgrounds \cite{BertiCardosoStarinets_2009_review,TimeEvolutionQuasiSchwarzschild_EPJC}. These perturbative analyses serve as direct physical probes of strong gravity, while simultaneously offering theoretically exact test systems whose field equations can be derived and analyzed without approximation within the linear regime.

\vspace{0.05cm}
\setlength{\parindent}{0pt}

The study of electromagnetic perturbations of black holes has a long history, and many of their fundamental properties are well established \cite{thorne1982electrodynamics,mashhoon1973scattering,hollowood2009refractive,moncrief1974gravitational,regge1957stability}. In particular, it is known that in four-dimensional static, spherically symmetric spacetimes the Maxwell field perturbations can be decomposed into axial (odd-parity) and polar (even-parity) sectors, and after elimination of gauge degrees of freedom both sectors satisfy decoupled Schr\"odinger-like radial wave equations with identical spectra for the pure electromagnetic field in the linear Maxwell case around Schwarzschild and Reissner-Nordstr\"om backgrounds~\cite{chandrasekhar1983mathematical}. Traditional treatments of black hole perturbations rely on auxiliary coordinate transformations, most notably the introduction of the tortoise coordinate $r_*$ defined through $dr_*/dr=f(r)^{-1}$, which maps the event horizon to $r_* \to -\infty$ and simplifies the radial dynamics and the identification of effective potentials~\cite{regge1957stability,zerilli1970effective}. While technically effective, such transformations tend to obscure the direct geometric origin of several key features of wave propagation, including gravitational redshift, near-horizon behavior, and the physical interpretation of the background potentials. Coordinate redefinitions can further blur the distinction between genuine physical effects associated with spacetime curvature and artifacts introduced by a particular parametrization of the radial coordinate. This is especially relevant when discussing optical analogies, dispersive behavior, or horizon related phenomena, where geometric transparency is essential for physical interpretation. These considerations motivate the development of formulations that remain entirely within Schwarzschild-like coordinates and preserve a direct correspondence between wave dynamics and spacetime geometry.

\vspace{0.05cm}
\setlength{\parindent}{0pt}

In this work, we study electromagnetic wave propagation in static, spherically symmetric black hole spacetimes that proceeds directly from the covariant Maxwell equations and remains entirely within Schwarzschild-like coordinates. No auxiliary variables, coordinate transformations, or horizon regular constructions are introduced at any stage. This approach ensures that every structure appearing in the radial wave equation admits a clear geometric interpretation and that the role of spacetime curvature is explicitly identifiable throughout the analysis. A central methodological principle of the present work is the systematic enforcement of gauge invariance. Electromagnetic perturbations contain inherent gauge freedom, and a physically meaningful description requires a clean separation between true dynamical degrees of freedom and gauge artifacts \cite{Molina,Ruffini}. By performing a complete parity decomposition of the electromagnetic four potential and explicitly eliminating gauge dependent components, we recover a single gauge invariant master variable that captures the full physical content of the electromagnetic field. As expected from general properties of Maxwell theory, both axial and polar sectors reduce to the same radial equation, and this well known isospectrality emerges naturally within the covariant framework as a consistency property rather than a separate assumption \cite{Molina,Ruffini}. \textcolor{black}{It should be emphasized that the parity decomposition of electromagnetic perturbations and the resulting gauge-invariant master equations are well-established results in black-hole perturbation theory \cite{chandrasekhar1983mathematical,Molina,Ruffini}. In the present work, these known constructions are adopted as a starting point for developing an alternative optical interpretation based on an effective refractive-index formulation derived directly in Schwarzschild coordinates.}

\vspace{0.05cm}
\setlength{\parindent}{0pt}

Beyond its formal clarity, the resulting master equation admits a natural wave optical interpretation. By eliminating first derivative terms through an appropriate field redefinition, the radial dynamics can be cast into a Helmholtz type equation. This reformulation motivates the introduction of an effective refractive index that depends explicitly on position and frequency and encodes gravitational redshift, curvature effects, and angular momentum within a single optical quantity \cite{o1,o2,o3,o4,o5}. Unlike heuristic analogies, this refractive index arises directly from the exact Maxwell equations and therefore possesses precise mathematical and physical meaning. The primary focus of this paper is on static, spherically symmetric black hole spacetimes, with particular emphasis on the Schwarzschild geometry as the simplest and most transparent example \cite{s1,s2,s3,s4,s5,s6,s7,s8,s9,s10,photon-sphere}. Nevertheless, the formalism developed here applies to a broad class of static black hole solutions satisfying mild geometric conditions. By maintaining manifest gauge invariance and avoiding coordinate artifacts, the resulting framework provides a robust foundation for analytical, semiclassical, and numerical studies of electromagnetic wave propagation in strong gravitational fields.

\vspace{0.05cm}
\setlength{\parindent}{0pt}

\textcolor{black}{Additional studies of electromagnetic fields in black-hole spacetimes include investigations of quantum aspects of electromagnetic fields and their implications for black-hole thermodynamics and low-energy phenomena \cite{33,34}. Furthermore, optical approaches to gravity have long played an important role in understanding wave propagation in curved spacetime and gravitational lensing phenomena \cite{35}.}

\vspace{0.05cm}
\setlength{\parindent}{0pt}

The structure of the paper is as follows. In Section~\ref{sec:EM-settings}, we formulate source free Maxwell dynamics on a general static, spherically symmetric black hole background and perform a complete parity decomposition of the electromagnetic field. Gauge invariant variables are constructed explicitly, and the reduction to a single master equation is demonstrated. In Section~\ref{sec:Schwarzschild}, the general formalism is specialized to the Schwarzschild geometry, where the parity independent radial equation is expressed entirely in Schwarzschild coordinates. Section~\ref{sec:refractive-index} recasts the exact radial equation into a Helmholtz type form and introduces the effective refractive index, providing a unified optical interpretation of gravitational redshift, curvature induced scattering, and turning point structure. Finally, Section~\ref{sec:conc} summarizes the main results and discusses their physical implications and possible extensions.

\section{Electromagnetic Wave Propagation: preliminary settings}\label{sec:EM-settings}

The dynamics of electromagnetic fields in curved spacetime offers a transparent and mathematically controlled framework for studying the relationship between spacetime geometry and field theory. \textcolor{black}{The decomposition of Maxwell perturbations into axial and polar sectors presented below follows the standard gauge-invariant formalism developed in previous studies \cite{chandrasekhar1983mathematical,Molina,Ruffini}. These established results are summarized here to provide a self-contained derivation of the master equation that forms the basis for the refractive-index analysis introduced in subsequent sections.} In the context of black hole spacetimes, electromagnetic perturbations play a dual role. On the one hand, they serve as physically observable probes of the underlying geometry through scattering processes and quasinormal ringing. On the other hand, from a theoretical standpoint, they constitute a linear, gauge-invariant test field whose equations of motion can be derived exactly and analyzed without ambiguity. These features render Maxwell theory an ideal laboratory for investigating wave propagation, stability properties, and mode decomposition in curved backgrounds. In a general curved spacetime endowed with a metric $g_{\mu\nu}$ and the associated Levi-Civita connection $\nabla_\mu$, the source-free Maxwell equations can be written in covariant form as \cite{Molina,Ruffini}
\begin{equation}
\nabla_\mu F^{\mu\nu}
=\frac{1}{\sqrt{-g}}\partial_\mu\!\left(\sqrt{-g}\,F^{\mu\nu}\right)=0,
\qquad
F_{\mu\nu}=\partial_\mu A_\nu-\partial_\nu A_\mu,
\label{2.1}
\end{equation}
where $A_\mu$ denotes the electromagnetic four-potential and $F_{\mu\nu}$ is the antisymmetric field-strength tensor. The definition of $F_{\mu\nu}$ in terms of the potential automatically implies the Bianchi identity $\nabla_{[\lambda}F_{\mu\nu]}=0$, while gauge invariance under the transformation $A_\mu \rightarrow A_\mu + \partial_\mu \Lambda$ guarantees that only two physical, transverse polarizations propagate \cite{Molina,Ruffini}. In the present methodological framework, we focus on a static, spherically symmetric spacetime described in standard Schwarzschild-like coordinates by the line element \cite{chandrasekhar1983mathematical,regge1957stability,Fabbri1975}
\begin{equation}
ds^2=-f(r)\,dt^2+h(r)\,dr^2+r^2\left(d\theta^2+\sin^2\theta\,d\phi^2\right),
\label{2.2}
\end{equation}
where we restrict attention to a physically relevant class of black hole geometries for which
\(
h(r)=\frac{1}{f(r)}.
\)
This condition is satisfied by Schwarzschild, Reissner-Nordstr\"om, and many other static black hole solutions, and it ensures that the radial coordinate $r$ coincides with the areal radius of the two-spheres of symmetry. With this identification, the metric takes the simplified form
\begin{equation}
ds^2=-f(r)\,dt^2+\frac{dr^2}{f(r)}+r^2\left(d\theta^2+\sin^2\theta\,d\phi^2\right),
\label{2.3}
\end{equation}
with inverse metric
\begin{equation}
g^{\mu\nu}=\mathrm{diag}\!\left(-\frac{1}{f(r)},\,f(r),\,\frac{1}{r^2},\,\frac{1}{r^2\sin^2\theta}\right),
\label{2.4}
\end{equation}
and determinant
\begin{equation}
g=-r^4\sin^2\theta,
\qquad
\sqrt{-g}=r^2\sin\theta.
\label{2.5}
\end{equation}
It is worth noting that, for this class of metrics, the determinant is independent of the specific form of the lapse function $f(r)$. The spherical symmetry of the background spacetime allows a complete decomposition of electromagnetic perturbations according to their behaviour under parity transformations \cite{Molina,Ruffini}
\(
(\theta,\phi)\rightarrow(\pi-\theta,\phi+\pi).
\)
Under this transformation, fields may either remain invariant (even, or polar parity) or change sign (odd, or axial parity). Correspondingly, electromagnetic perturbations naturally split into two independent parity sectors, which decouple identically at the linearized level and can therefore be analyzed separately. Accordingly, the electromagnetic four-potential $A_\mu=(A_t,A_r,A_\theta,A_\phi)$ admits two distinct decompositions. The first corresponds to the axial (odd-parity) sector and is given by \cite{Molina,Ruffini}
\begin{equation}
\begin{split}
        &A_t=0, \qquad A_r=0, \\
        &A_\theta=\frac{\psi(r,t)}{\sin\theta}\,\partial_\phi Y^{\ell m}(\theta,\phi), \\
        &A_\phi=-\psi(r,t)\sin\theta\,\partial_\theta Y^{\ell m}(\theta,\phi),
\end{split}
\label{2.6}
\end{equation}
where $\psi(r,t)$ encodes the dynamical degrees of freedom of the axial electromagnetic perturbations. The second decomposition corresponds to the polar (even-parity) sector and takes the form \cite{Molina,Ruffini}
\begin{equation}
\begin{split}
            &A_t=Q_{1}(r,t)\,Y^{\ell m}(\theta,\phi), \qquad
            A_r=Q_{2}(r,t)\,Y^{\ell m}(\theta,\phi), \\
            &A_\theta=K(r,t)\,\partial_\theta Y^{\ell m}(\theta,\phi), \qquad
            A_\phi=K(r,t)\,\partial_\phi Y^{\ell m}(\theta,\phi),
\end{split}
\label{2.7}
\end{equation}
where $Q_{1}(r,t)$, $Q_{2}(r,t)$, and $K(r,t)$ characterize the polar electromagnetic perturbations, and $Y^{\ell m}(\theta,\phi)$ denote the scalar spherical harmonics. In the following subsections, we explicitly demonstrate that both parity sectors reduce to identical master equations governing electromagnetic wave propagation on the black hole background. \textcolor{black}{The parity decompositions employed above follow the standard treatment of electromagnetic perturbations in static, spherically symmetric spacetimes developed in Refs. \cite{Molina,Ruffini}. They are summarized here to establish the notation and provide a self-contained derivation of the master equation used in the subsequent refractive-index formulation.}

\subsection{Electromagnetic wave propagation in the axial sector}

In the axial (odd-parity) sector, electromagnetic perturbations are described by the vector potential $A_\mu$ given in Eq.~(\ref{2.6}). Substituting this ansatz into the source-free Maxwell equations (\ref{2.1}), and making explicit use of the background metric (\ref{2.3}) together with the determinant (\ref{2.5}), we obtain the following set of field equations (see also \cite{Molina,Ruffini}):
\begin{equation}
\begin{split}
    &\partial_\mu \sqrt{-g}F^{\mu t}=0
    \Rightarrow
    \partial_\theta \sin\theta \,\partial _tA_\theta
    +\partial_\phi \frac{1}{\sin\theta}\partial_t A_\phi=0 ,\\[0.3em]
    &\partial_\mu \sqrt{-g}F^{\mu r}=0
    \Rightarrow
    \partial_\theta \sin\theta\,\partial_rA_\theta
    +\partial_\phi \frac{1}{\sin\theta}\partial_r A_\phi=0,\\[0.3em]
    &\partial_\mu \sqrt{-g}F^{\mu \theta}=0
    \\[0.2em]
    &\Rightarrow
    -\frac{1}{f(r)}\partial_t^2 A_\theta
    +\partial_r f(r)\,\partial_rA_\phi
    +\frac{1}{r^2\sin^2\theta}
    \partial_\phi(\partial_\phi A_\theta-\partial_\theta A_\phi)=0, \\[0.3em]
    &\partial_\mu \sqrt{-g}F^{\mu \phi}=0
    \\[0.2em]
    &\Rightarrow
    -\frac{1}{f(r)}\partial_t^2 A_\phi
    +\partial_r f(r) \partial_r A_\phi
    +\sin\theta \,\partial_\theta
    \frac{1}{r^2\sin\theta}
    (\partial_\theta A_\phi-\partial_\phi A_\theta)=0.
\end{split}
\label{2.8}
\end{equation}
A straightforward substitution of the axial ansatz (\ref{2.6}) into the first two equations above reveals that both reduce identically to the angular identity
\[
\partial_\theta\partial_\phi Y^{\ell m}
-\partial_\phi\partial_\theta Y^{\ell m}=0,
\]
which is trivially satisfied due to the smoothness of the scalar spherical harmonics. Consequently, the temporal and radial Maxwell equations impose no additional dynamical constraints in the axial sector. The remaining two equations in (\ref{2.8}) are likewise identical and combine to yield a single second-order partial differential equation governing the axial master variable $\psi(r,t)$,
\begin{equation}
  -\partial_t^2\psi(r,t)
  +\partial_r\!\left(f(r)\partial_r\psi(r,t)\right)
  -\frac{\ell(\ell+1)}{r^2}\psi(r,t)=0.
\label{2.9}
\end{equation}
This equation describes the propagation of electromagnetic waves on the curved background and already exhibits the characteristic centrifugal barrier associated with angular momentum $\ell$. Assuming a harmonic time dependence of the form
\(
\psi(r,t)=e^{-i\omega t}\psi(r),
\)
Eq.~(\ref{2.9}) reduces to the radial equation
\begin{equation}
    \left(
    \partial_r^2
    +\frac{f'(r)}{f(r)}\partial_r
    +\frac{\omega^2}{f(r)^2}
    -\frac{\ell(\ell+1)}{r^2 f(r)}
    \right)\psi(r)=0.
\label{2.10}
\end{equation}
It is convenient to recast this equation into a Schr\"odinger- or Helmholtz-like form by eliminating the first-derivative term through the field redefinition
\(
\psi(r)=\chi(r)/\sqrt{f(r)}.
\)
This transformation leads to
\begin{equation}
\begin{split}
    &\chi''(r)+k^2(r,\omega)\chi(r)=0,\\[0.2em]
    &k^2(r,\omega)=
    \frac{\omega^2}{f(r)^2}
    -\frac{\ell(\ell+1)}{r^2f(r)}
    +\frac{f'(r)^2}{4f(r)^2}
    -\frac{f''(r)}{2f(r)}.
\end{split}
\label{2.11}
\end{equation}
This equation explicitly displays the local wave number governing axial electromagnetic perturbations, fully encoding the influence of spacetime curvature through the metric function $f(r)$.

\subsection{Electromagnetic wave propagation in the polar sector}

We now turn to the polar (even-parity) sector \cite{Molina,Ruffini}, described by the vector potential ansatz given in Eq.~(\ref{2.7}). Substitution of this decomposition into the Maxwell equations (\ref{2.1}) yields the coupled system
\begin{equation}
    \begin{split}
        &\partial_\mu \sqrt{-g}F^{\mu t}=0 \Rightarrow \\
        &\partial_r r^2(\partial_rQ_1-\partial_tQ_2)
        +\frac{\ell(\ell+1)}{f(r)}\left[\partial_t K-Q_1\right]=0,\\[0.3em]
        &\partial_\mu \sqrt{-g}F^{\mu r}=0 \Rightarrow \\
        &-\partial_t\left(\partial_t Q_2-\partial_r Q_1\right)
        +\frac{f(r)\ell(\ell+1)}{r^2}\left[\partial_r K-Q_2\right]=0,\\[0.3em]
        &\partial_\mu \sqrt{-g}F^{\mu \theta}=0 \Rightarrow \\
        &\partial_t(Q_1-\partial_t K)
        +f(r)\partial_r f(r)(\partial_r K-Q_2)=0,\\[0.3em]
        &\partial_\mu \sqrt{-g}F^{\mu \phi}=0 \Rightarrow \\
        &\partial_t(Q_1-\partial_t K)
        +f(r)\partial_r f(r)(\partial_r K-Q_2)=0.
    \end{split}
\label{2.12}
\end{equation}
As expected from spherical symmetry, the last two equations are identical, reflecting the degeneracy between the $\theta$ and $\phi$ components. To decouple the system, we introduce the master variable
\[
b(r,t)
=-\frac{r^2}{\ell(\ell+1)}(\partial_r Q_1-\partial_t Q_2),
\]
which allows the first two equations in (\ref{2.12}) to be rewritten as
\begin{equation}
    \begin{split}
        f(r)\partial_rb(r,t)-[\partial_tK-Q_1]=0,\\
        \partial_tb(r,t)-f(r)[\partial_rK-Q_2]=0.
    \end{split}
\label{2.13}
\end{equation}
Combining these relations and assuming harmonic time dependence
\(
b(r,t)=e^{-i\omega t}b(r),
\)
we arrive at the radial equation
\begin{equation}
    \partial_rf(r)\partial_r b(r)
    +\left[
    \frac{\omega^2}{f(r)}
    +\frac{\ell(\ell+1)}{r^2}
    \right] b(r)=0.
\label{2.14}
\end{equation}
Finally, performing the same field redefinition as in the axial sector,
\(
b(r)=\chi(r)/\sqrt{f(r)},
\)
one finds that the polar perturbations satisfy exactly the same Helmholtz-like equation (\ref{2.11}), with an identical effective potential.

\setlength{\parindent}{0pt}
\vspace{0.05cm}

The complete equivalence of the axial and polar sectors constitutes a central and physically significant result. Despite their distinct parity properties and tensorial structures at the level of the vector potential, both sectors reduce to the same master wave equation governed by the local wave number (\ref{2.11}). This is a direct manifestation of the underlying gauge invariance and conformal properties of Maxwell theory in four-dimensional curved spacetime. Consequently, electromagnetic perturbations on static, spherically symmetric black hole backgrounds propagate without parity-dependent splitting, in sharp contrast to the gravitational case where axial and polar modes obey inequivalent dynamics. The universal form of the potential further implies that scattering properties, quasinormal spectra, and stability features of electromagnetic waves are entirely determined by the background geometry encoded in $f(r)$, independently of the parity sector under consideration.

\section{Schwarzschild Black Hole}\label{sec:Schwarzschild}

\setlength{\parindent}{0pt}

We now specialize the general formalism for electromagnetic perturbations developed in the previous sections to the Schwarzschild black hole. As the simplest static, spherically symmetric, asymptotically flat vacuum solution of Einstein’s equations, the Schwarzschild geometry provides an ideal setting in which the essential features of electromagnetic wave propagation in curved spacetime can be analyzed in a transparent and controlled manner. In standard Schwarzschild coordinates, the spacetime metric is given by Eq. \eqref{2.3} with
\begin{equation}
f(r)=1-\frac{2M}{r},
\end{equation}
where \(M\) denotes the mass of the black hole. The coordinate singularity at \(F(r_h)=0\Rightarrow r_{h}=2M\) corresponds to the event horizon, while the region \(r>2M\) describes the exterior spacetime accessible to static observers \cite{s1,s2,s3,s4,s5,s6,s7,s8,s9,s10}. Throughout this section, we restrict attention to this exterior region and work entirely in Schwarzschild coordinates, without introducing horizon-regular variables. As shown previously, electromagnetic perturbations on a static, spherically symmetric background admit a decomposition into odd- and even-parity sectors. Despite their distinct angular structures and gauge properties, both sectors reduce (after elimination of gauge redundancy) to the same gauge-invariant master equation. Consequently, the physical electromagnetic degrees of freedom are fully described by a single scalar master variable \(\Psi(t,r)\), independent of parity. For spacetimes satisfying \(h(r)=1/f(r)\), the master equation governing electromagnetic perturbations takes the universal form \eqref{2.11} with \(\ell=1,2,3,\dots\). This equation holds identically for both axial and polar perturbations and reflects the exact isospectrality of the electromagnetic field in the Schwarzschild geometry.

\vspace{0.05cm}
\setlength{\parindent}{0pt}

The master radial equation can be re-expressed in the following form
\begin{equation}
\chi''(r)+\left[\frac{\omega^2}{f(r)^2}-V_{\rm EM}(r)\right]\chi(r)=0,
\end{equation}
where
\begin{equation}
V_{\rm EM}(r)=\frac{\ell(\ell+1)}{r^2 f(r)}-\frac{(f'(r))^2}{4 f(r)^2}+\frac{f''(r)}{2 f(r)}.
\end{equation}
For the Schwarzschild metric, the derivatives of the metric function are \(f'(r)=\frac{2M}{r^2}\) and \(f''(r)=-\frac{4M}{r^3}\). Substitution into the general expression for \(V_{\rm EM}(r)\) gives
\begin{equation}
\begin{split}
&V_{\rm EM}^{\rm Schw}(r)
=
\frac{\ell(\ell+1)}{r^{2}\left(1-\frac{2M}{r}\right)}
-
\frac{M^{2}}{r^{4}\left(1-\frac{2M}{r}\right)^{2}}
-
\frac{2M}{r^{3}\left(1-\frac{2M}{r}\right)}.\\
&V_{\mathrm{EM}}^{\mathrm{Schw}}(r)\Rightarrow \frac{r(r-2M)\,\ell(\ell+1)}{r^{2}(r-2M)^{2}}
-
\frac{M^{2}}{r^{2}(r-2M)^{2}}
-
\frac{2M}{r^{2}(r-2M)}\label{eff-pot}
\end{split}
\end{equation}
This potential governs the radial behavior of all electromagnetic perturbations (independent of parity) in the Schwarzschild background when the wave equation is expressed directly in terms of the Schwarzschild radial coordinate. The derivative-dependent terms arise solely from the field redefinition used to remove first derivatives and reflect the influence of spacetime curvature on the normalization of the radial wave function rather than additional physical interactions. In the asymptotic region \(r\to\infty\), the spacetime approaches flat Minkowski space and the effective potential reduces to
\begin{equation}
V_{\rm EM}^{\rm Schw}(r)\sim \frac{\ell(\ell+1)}{r^2},
\end{equation}
corresponding to the familiar centrifugal barrier for electromagnetic waves in flat spacetime. This behavior ensures the correct matching to plane-wave solutions at spatial infinity and underlies the standard formulation of scattering theory. As \(r\to 2M\), the effective potential diverges due to factors of \(1/f(r)\). This divergence reflects the coordinate singularity of the Schwarzschild radial coordinate rather than any physical pathology of the electromagnetic field. Within this coordinate system, it enforces the causal structure of the event horizon by suppressing outgoing modes originating from \(r=2M\). Between the horizon and infinity, the effective potential forms a curvature-induced barrier whose height and shape depend on the angular momentum number \(\ell\). Higher multipole modes experience stronger reflection, while lower multipoles are more readily absorbed by the black hole. This potential therefore controls the scattering properties, absorption cross sections, and quasinormal-mode spectrum of electromagnetic perturbations in the Schwarzschild spacetime (see also \cite{CardosoLemos_PRD64_2001,Konoplya2005}).

\section{Effective Refractive Index}
\label{sec:refractive-index}

\setlength{\parindent}{0pt}

A powerful and physically intuitive interpretation of electromagnetic wave propagation in curved spacetime can be obtained by recasting the radial wave equation into a form analogous to wave propagation in an inhomogeneous optical medium. In this optical analogy, the influence of spacetime curvature is encoded in an effective, position- and frequency-dependent refractive index that governs the local radial wavenumber of the field. This perspective provides a unified description of gravitational redshift, curvature-induced scattering, and angular-momentum barriers within a single framework. The starting point of the construction is the radial equation obtained after separation of variables and removal of first-derivative terms by an appropriate field redefinition. It should be noted that it is structurally equivalent to a one-dimensional Helmholtz equation (with $c=1$) of the form \cite{o1,o2,o3,o4,o5}
\begin{equation}
\chi''(r)
+
\omega^2\,n^2(r,\omega)\,\chi(r)
=
0,
\end{equation}
where \(\chi(r)\) is the redefined radial amplitude and \(\omega\) is the frequency of the electromagnetic wave. This formal correspondence motivates the definition of an effective refractive index \(n(r,\omega)\), which captures the cumulative effect of spacetime geometry and angular momentum on the radial propagation of electromagnetic waves. By direct comparison with the radial master equation, the effective refractive index is identified as
\begin{equation}
n^2(r,\omega)
=
\frac{1}{f(r)^2}
-
\frac{V_{\rm EM}(r)}{\omega^2},
\end{equation}
where \(f(r)\) is the metric function appearing in the Schwarzschild line element and \(V_{\rm EM}(r)\) is the unique, parity-independent effective potential governing electromagnetic perturbations. The explicit dependence of \(n(r,\omega)\) on both position and frequency implies that wave propagation in curved spacetime is intrinsically dispersive when viewed from the perspective of a distant observer. Specializing to the Schwarzschild geometry, for which \(f(r)=1-\frac{2M}{r}\), and using the explicit form of the electromagnetic effective potential derived previously, the refractive index squared becomes
\begin{equation}
\begin{split}
n_{\rm Schw}^2(r,\omega)
&=\frac{1}{\left(1-\frac{2M}{r}\right)^2}\\
&\quad-\frac{1}{\omega^2}
\left[
\frac{\ell(\ell+1)}{r^2\left(1-\frac{2M}{r}\right)}
-\frac{M^2}{r^4\left(1-\frac{2M}{r}\right)^2}
-\frac{2M}{r^3\left(1-\frac{2M}{r}\right)}
\right].
\label{ref-index}
\end{split}
\end{equation}
The first term originates purely from the spacetime geometry and encodes the gravitational redshift experienced by electromagnetic waves as measured by an asymptotic observer. To quantify its effect on propagation, consider radial transmission from an emission point \(r_0>2M\) to a distant observer located at \(R\gg2M\). In the near-horizon region, writing \(r=2M+\epsilon\) with \(\epsilon\ll2M\), one finds \(f(r)\simeq\epsilon/(2M)\) and \(f(r)^{-2}\simeq4M^2/\epsilon^2\), demonstrating that the metric contribution to \(n^2\) diverges quadratically as \(\epsilon^{-2}\), while the curvature-induced terms scale more weakly and are suppressed by the factor \(1/\omega^2\). As a result, the refractive index exhibits the universal near-horizon behavior \(n(r,\omega)\simeq1/f(r)\), independent of angular momentum and frequency. The corresponding optical path length for radial propagation,
\begin{equation}
\mathcal{L}_{\rm opt}(r_0\to R)
=
\int_{r_0}^{R} n(r,\omega)\,dr,
\end{equation}
is therefore governed near the horizon by
\begin{equation}
\mathcal{L}_{\rm opt}(r_0\to R)
\simeq
\int_{r_0}^{R}\frac{dr}{1-\frac{2M}{r}}
=
(R-r_0)
+
2M\ln\!\left(\frac{R-2M}{r_0-2M}\right).
\end{equation}
As the emission point approaches the horizon, \(r_0\to2M^+\), the logarithmic term diverges, implying an infinite optical path length. This divergence is purely geometric and reflects the infinite Schwarzschild coordinate time required for signals emitted arbitrarily close to the horizon to reach distant observers. The remaining terms in Eq.~\eqref{ref-index}, originating from the curvature-induced effective potential, encode angular-momentum-dependent scattering, frequency-dependent dispersion, and the emergence of classically forbidden radial regions, while remaining subdominant in the near-horizon limit. In the asymptotic region \(r\to\infty\), the Schwarzschild spacetime approaches flat Minkowski space and the effective potential decays rapidly. Consequently, the refractive index approaches unity,
\begin{equation}
n_{\rm Schw}(r,\omega)
\to
1
-
\frac{\ell(\ell+1)}{2\omega^2 r^2}
+
\mathcal{O}\!\left(\frac{1}{r^3}\right),
\end{equation}
indicating that electromagnetic waves propagate freely at large distances, with only small corrections due to angular momentum. This behavior ensures consistency with standard flat-space electrodynamics and confirms the asymptotic flatness of the Schwarzschild geometry within the optical analogy. At intermediate radii, the sign of \(n_{\rm Schw}^2(r,\omega)\) determines the qualitative nature of wave propagation. In regions where
\begin{equation}
\omega^2 > V_{\rm EM}(r),
\end{equation}
the refractive index is real and the radial solutions are oscillatory, corresponding to propagating electromagnetic waves. In contrast, when
\begin{equation}
\omega^2 < V_{\rm EM}(r),
\end{equation}
the refractive index becomes imaginary and the solutions decay or grow exponentially. These evanescent regions correspond to classically forbidden zones in the WKB sense and are responsible for partial reflection of waves by the curvature-induced potential barrier. In the low-frequency regime, where \(\omega^2\ll V_{\rm EM}(r)\) over an extended radial range, the refractive index is predominantly imaginary,
\begin{equation}
n_{\rm Schw}(r,\omega)
\approx
i\,\sqrt{\frac{V_{\rm EM}(r)}{\omega^2}},
\end{equation}
indicating strong suppression of wave propagation. Physically, low-frequency electromagnetic waves are efficiently reflected by the angular-momentum barrier and are unable to penetrate deeply toward the black hole. This behavior underlies the characteristic suppression of low-energy absorption cross sections. In the opposite, high-frequency limit \(\omega\to\infty\), the contribution of the effective potential becomes negligible and the refractive index reduces to
\begin{equation}
n_{\rm Schw}(r,\omega\to\infty)
\approx
\frac{1}{1-\frac{2M}{r}},
\end{equation}
corresponding to the geometric-optics regime. In this limit, electromagnetic waves follow null geodesics of the Schwarzschild spacetime and become insensitive to the detailed structure of the potential barrier. The refractive index depends solely on the gravitational redshift factor and is independent of angular momentum. The turning points of radial motion are determined by the condition \(n_{\rm Schw}^2(r_t,\omega)=0\), which defines the boundaries between oscillatory and evanescent regions. For typical values of \(\ell\) and \(\omega\), this condition admits two real solutions: an inner turning point located near the event horizon and an outer turning point associated with the peak of the effective potential. The latter lies close to the photon sphere at \(r=3M\), where unstable circular null orbits play a central role in wave scattering. Near a turning point \(r_t\), the refractive index admits a linear expansion,
\begin{equation}
n_{\rm Schw}^2(r,\omega)
\approx
\left.
\frac{d}{dr}n_{\rm Schw}^2(r,\omega)
\right|_{r=r_t}
(r-r_t),
\end{equation}
which governs the local transition between propagating and evanescent behavior. This structure controls reflection and transmission coefficients and determines tunneling probabilities through the curvature-induced barrier. The effective refractive index formulation thus provides a comprehensive and unified description of electromagnetic wave propagation in Schwarzschild spacetime. It incorporates gravitational redshift, angular-momentum scattering, frequency-dependent dispersion, and the existence of classically forbidden regions within a single optical framework. Beyond its conceptual clarity, this approach offers practical advantages for WKB analyses, semiclassical approximations, and numerical computations of scattering amplitudes and quasinormal-mode spectra in black-hole spacetimes.

\begin{figure*}[ht]
\centering
\includegraphics[scale=0.60]{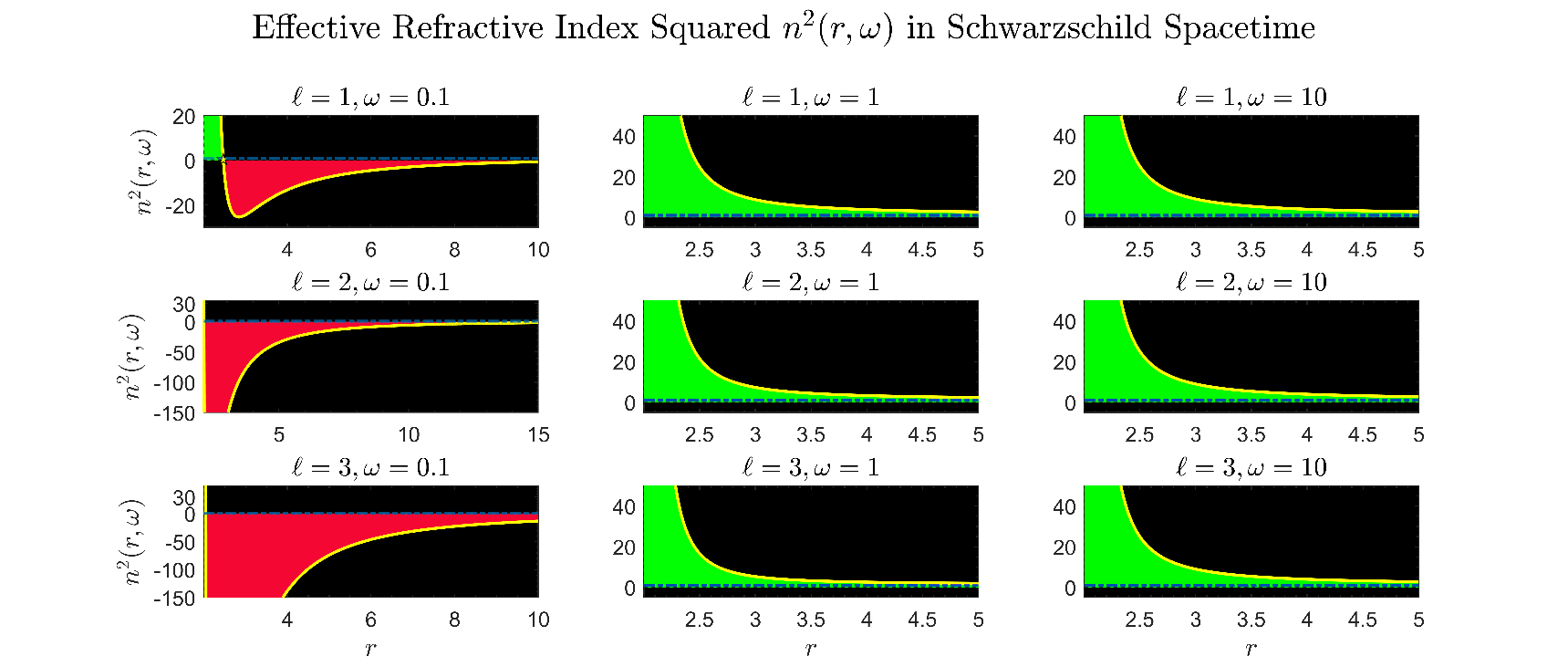}\\
\caption{
\fontsize{7.6}{8.8}\selectfont
Effective refractive index squared $n^2(r,\omega)$ for electromagnetic waves propagating in a Schwarzschild spacetime with mass $M=1$. Nine subplots correspond to combinations of angular momentum numbers $\ell = 1,2,3$ (rows) and wave frequencies $\omega = 0.1, 1, 10$ (columns). The radial coordinate extends from just outside the horizon $r_{\rm min}=2.0001\,M$ to $r_{\rm max}=10\,M$. In each subplot, the colored curves represent $n^2(r,\omega)$, red shaded regions indicate $n^2<0$, and green shaded regions indicate $n^2>0$. The blue dot-dashed line corresponds to $n^2=1$ as a reference to the asymptotic vacuum limit. The black dashed line indicates the event horizon at $r=2M$.}
\label{fig:ref-index-SCHW}
\end{figure*}
The effective refractive index squared $n^2(r,\omega)$ (see Figure \ref{fig:ref-index-SCHW}) reveals the profound dispersive and spatially varying nature of electromagnetic wave propagation in a Schwarzschild gravitational field. For relatively small frequencies ($\omega = 0.1$), $n^2(r,\omega)$ exhibits extensive negative regions near the horizon, corresponding to classically forbidden radial propagation and purely evanescent behavior. These negative-$n^2(r,\omega)$ zones shrink and move closer to the horizon as the wave frequency increases ($\omega = 1, 10$), reflecting the diminishing influence of spacetime curvature on higher-energy photons. The angular momentum barrier, determined by $\ell$, introduces additional structure in $n^2(r,\omega)$: higher $\ell$ enhances the centrifugal contribution to the effective potential, producing more pronounced positive $n^2(r,\omega)$ peaks at intermediate radii. All curves converge to $n^2(r,\omega) \to 1$ at large $r$, consistent with asymptotic flat spacetime. The combined effects of $\omega$ and $\ell$ demonstrate the dual influence of curvature and angular momentum on radial propagation: low-frequency, high-$\ell$ modes are more strongly confined near the black hole, whereas high-frequency, low-$\ell$ modes propagate almost freely throughout the exterior spacetime. Overall, $n^2(r,\omega)$ provides a unified description of gravitationally induced optical phenomena, encompassing dispersion, evanescence, and angular-momentum barriers within a single refractive-index framework.

\section{Summary and Conclusions}
\label{sec:conc}

\setlength{\parindent}{0pt}

In this work, a complete and geometrically transparent description of electromagnetic wave propagation in static, spherically symmetric black hole spacetimes has been studied. The analysis is carried out directly at the level of the covariant Maxwell equations and remains entirely within Schwarzschild-like coordinates throughout. By avoiding auxiliary coordinate transformations and horizon regular variables, all features of the wave dynamics retain a clear and direct interpretation in terms of the underlying spacetime geometry.

\vspace{0.05cm}
\setlength{\parindent}{0pt}

A systematic parity decomposition of the electromagnetic four potential is performed, and gauge dependent components are explicitly eliminated. As expected from well established results in black hole perturbation theory, axial and polar electromagnetic perturbations are found to satisfy the same radial master equation. This exact isospectrality is not a new physical result, but rather a known property of Maxwell fields in four dimensional static spacetimes. In the present framework, however, it emerges naturally and without additional assumptions, serving as a nontrivial consistency check of the fully covariant and gauge invariant construction. The physical content of electromagnetic perturbations is therefore shown to be completely captured by a single scalar master variable, independent of parity. The resulting master equation takes the form of a second order radial wave equation whose coefficients depend exclusively on the background metric function and the angular momentum number. When the condition that the radial and temporal metric functions are reciprocal is imposed, the equation admits a Schrödinger type representation written entirely in terms of the areal radius. The associated effective potential is uniquely determined by the spacetime geometry and contains contributions arising from angular momentum, spacetime curvature, and normalization effects induced by the radial dependence of the metric.

\vspace{0.05cm}
\setlength{\parindent}{0pt}

Specializing to the Schwarzschild geometry, the local wave number is obtained in closed analytical form. In the asymptotic region, it reduces smoothly to the familiar centrifugal barrier of flat space electrodynamics, ensuring correct matching to plane wave solutions and consistency with standard scattering theory. Near the event horizon, it diverges in Schwarzschild coordinates, reflecting the causal structure of the horizon rather than any physical divergence of the electromagnetic field. This behavior enforces the suppression of outgoing modes originating from the horizon as viewed by distant observers.

\vspace{0.05cm}
\setlength{\parindent}{0pt}

\textcolor{black}{While the gauge-invariant treatment of electromagnetic perturbations and the associated parity-independent master equations are established results in the literature, the central contribution of this work is the reformulation of the resulting wave equation in terms of an effective refractive index defined directly in Schwarzschild coordinates.} By recasting the exact master equation into a Helmholtz type form, the influence of spacetime curvature on electromagnetic propagation is encoded in a position and frequency dependent optical response. This refractive index is derived directly from the Maxwell equations and provides an optical interpretation of curved spacetime electrodynamics.

\vspace{0.05cm}
\setlength{\parindent}{0pt}

In the Schwarzschild spacetime, the refractive index exhibits universal divergence near the event horizon, independent of frequency and angular momentum. This behavior reflects extreme gravitational redshift and corresponds to an infinite optical path length for signals approaching the horizon when measured in Schwarzschild time. At large radial distances, the refractive index approaches unity with corrections suppressed by inverse powers of the radius, demonstrating the recovery of flat space electrodynamics and optical transparency in the weak field regime.

\vspace{0.05cm}
\setlength{\parindent}{0pt}

At intermediate radii, the sign of the refractive index squared determines whether electromagnetic waves propagate or decay exponentially. Regions of negative refractive index squared correspond to evanescent behavior and define classically forbidden zones that control reflection and tunneling phenomena. The turning points separating these regions are closely associated with the curvature induced potential barrier and are typically located near the photon sphere, highlighting the connection between wave propagation and unstable null geodesics.

\vspace{0.05cm}
\setlength{\parindent}{0pt}

The frequency dependence of the refractive index reveals a smooth transition between wave dominated and geometric regimes. Low frequency modes experience strong attenuation due to extended evanescent regions, explaining the suppression of electromagnetic absorption at low energies. In the high frequency limit, curvature induced contributions become negligible and propagation approaches the geometric optics regime, where electromagnetic waves follow null geodesics of the background spacetime.

\vspace{0.05cm}
\setlength{\parindent}{0pt}

\textcolor{black}{An important direction for future work is the extension of the present refractive-index formulation to rotating black holes described by the Kerr geometry. Since astrophysical black holes are generally expected to possess angular momentum, understanding electromagnetic wave propagation in rotating backgrounds is of considerable physical interest. Such an extension would require incorporating the effects of axisymmetry, frame dragging, and the separability properties of the Teukolsky formalism. Investigating whether an analogous effective refractive-index description can be constructed in Kerr spacetime represents a promising avenue for future research.}

\vspace{0.05cm}
\setlength{\parindent}{0pt}

{\color{black}
This work provides a physically transparent framework for electromagnetic wave propagation in static black-hole spacetimes based on an effective refractive index description. While the covariant formulation of Maxwell theory and the corresponding electromagnetic mode decompositions in static gravitational backgrounds are well established, the present approach demonstrates how these known results can be systematically interpreted within an optical framework governed by an effective refractive index induced by spacetime curvature. This perspective unifies the description of gravitational redshift, angular momentum barriers, dispersion, and scattering processes, revealing them as manifestations of the underlying optical properties of the background spacetime. The resulting formalism retains the full consistency of the covariant theory while offering enhanced physical intuition and practical utility for semiclassical analyses, numerical studies, and generalizations to broader classes of spacetimes. Consequently, it provides a clear foundation for future investigations of wave-optical effects in strong gravitational fields.}

\begin{acknowledgements}
\textcolor{black}{We sincerely thank the reviewers for their careful evaluation of our manuscript and for their insightful comments and constructive suggestions. Their feedback has helped us improve the clarity, and overall quality of the manuscript. We also thank the Editor for handling our submission and providing us with the opportunity to revise our work.}
\end{acknowledgements}


\begin{thebibliography}{99}

{\fontsize{7.6}{8.6}\selectfont

\bibitem{thorne1982electrodynamics}
K. S. Thorne, D. Macdonald, Electrodynamics in curved spacetime: 3+1 formulation,
Mon. Not. R. Astron. Soc. {\bf 198}, 339--343 (1982),
\url{https://doi.org/10.1093/mnras/198.2.339}.

\bibitem{mashhoon1973scattering}
B. Mashhoon, Scattering of electromagnetic radiation from a black hole,
Phys. Rev.D {\bf 7}, 2807 (1973),
\url{https://doi.org/10.1103/PhysRevD.7.2807}.

\bibitem{hollowood2009refractive}
T. J. Hollowood, G. M.  Shore, R. J. Stanley, The refractive index of curved spacetime II: QED, Penrose limits and black holes,
JHEP {\bf 2009}, 089 (2009),
\url{https://doi.org/10.1088/1126-6708/2009/08/089}.

\bibitem{moncrief1974gravitational}
V. Moncrief, Gravitational perturbations of spherically symmetric systems. I. The exterior problem,
Ann. Phys. {\bf 88}, 323--342 (1974),
\url{https://doi.org/10.1016/0003-4916(74)90173-0}.

\bibitem{regge1957stability}
T. Regge, J. A. Wheeler, Stability of a Schwarzschild singularity, Phys. Rev. {\bf 108}, 1063 (1957),
\url{https://doi.org/10.1103/PhysRev.108.1063}.

\bibitem{BertiCardosoStarinets_2009_review}
E.~Berti, V.~Cardoso, and A.~O.~Starinets, Quasinormal modes of black holes and black branes, Class.\ Quantum Grav.\ {\bf 26}, 163001 (2009), \url{https://doi.org/10.1088/0264-9381/26/16/163001}.

\bibitem{CardosoLemos_PRD64_2001}
V.~Cardoso and J.~P.~S.~Lemos, Quasinormal modes of Schwarzschild–anti-de Sitter black holes: Electromagnetic and gravitational perturbations, Phys.\ Rev.\ D {\bf 64}, 084017 (2001), \url{https://doi.org/10.1103/PhysRevD.64.084017}.

\bibitem{Mamani2022_EPJC}
L.~A.~H.~Mamani, A.~D.~D.~Masa, L.~T.~Sanches, V. T. Zanchin,
Revisiting the quasinormal modes of the Schwarzschild black hole: Numerical analysis,
Eur.\ Phys.\ J.\ C {\bf 82}, 897 (2022),
\url{https://doi.org/10.1140/epjc/s10052-022-10865-1}.

\bibitem{Fabbri1975}
R.~Fabbri,
Scattering and absorption of electromagnetic waves by a Schwarzschild black hole,
Phys.\ Rev.\ D {\bf 12}, 933 (1975),
\url{https://doi.org/10.1103/PhysRevD.12.933}.

\bibitem{Konoplya2005}
R.~A.~Konoplya,
Massive vector field perturbations in the Schwarzschild background: Stability and quasinormal spectrum,
Phys.\ Rev.\ D {\bf 73}, 024009 (2006),
\url{https://doi.org/10.1103/PhysRevD.73.024009}.

\bibitem{TimeEvolutionQuasiSchwarzschild_EPJC}
O.~R.~de~Medeiros, M.~M.~Corr\^ea, and C.~F.~B.~Macedo,
Time evolution of perturbations in quasi-Schwarzschild black holes,
Eur.\ Phys.\ J.\ C {\bf 85}, 683 (2025),
\url{https://doi.org/10.1140/epjc/s10052-025-14422-4}.


\bibitem{chandrasekhar1983mathematical}
S. Chandrasekhar, The Mathematical Theory of Black Holes, Clarendon Press, Oxford (1983).

\bibitem{zerilli1970effective}
F. J. Zerilli, Effective potential for even-parity Regge-Wheeler gravitational perturbation equations,
Phys. Rev. Lett. {\bf 24}, 737–738 (1970), \url{https://doi.org/10.1103/PhysRevLett.24.737}.


\bibitem{Molina}
C. Molina, A. B. Pavan, and T. E. Medina Torrejón,
Electromagnetic perturbations in new brane world scenarios,
Phys.\ Rev.\ D {\bf 93}, 124068 (2016),
\url{https://doi.org/10.1103/PhysRevD.93.124068}.

\bibitem{Ruffini}
R. Ruffini, J. Tiomno, and C. V. Vishveshwara,
Electromagnetic field of a particle moving in a spherically symmetric black-hole background,
Lett.\ Nuovo Cim.\ {\bf 3}, 211--215 (1972),
\url{https://doi.org/10.1007/BF02772872}.


\bibitem{o1}
S. Gurtas Dogan, A. Guvendi, and O. Mustafa,
Geometric and wave optics in a BTZ optical metric-based wormhole,
Phys. Lett. B {\bf 868}, 139824 (2025),
\url{https://doi.org/10.1016/j.physletb.2025.139824}.

\bibitem{o2}
S. Gurtas Dogan, A. Guvendi, and O. Mustafa,
Ray and wave optics in an optical wormhole,
Phys. Lett. B {\bf 868}, 139626 (2025),
\url{https://doi.org/10.1016/j.physletb.2025.139626}.

\bibitem{o3}
S. Gurtas Dogan, A. Guvendi, and O. Mustafa,
Ray geodesics and wave propagation on the Beltrami surface: optics of an optical wormhole,
Eur. Phys. J. C {\bf 85}, 896 (2025),
\url{https://doi.org/10.1140/epjc/s10052-025-14644-6}.

\bibitem{o4}
F. Ahmed, A. Bouzenada, Geometric and Wave Optics in a Topologically Charged Perry-Mann type Wormhole with Disclinations,
Phys. Lett. B {\bf 868}, 139704 (2025),
\url{https://doi.org/10.1016/j.physletb.2025.139704}.

\bibitem{o5}
S. Gurtas Dogan, O. Mustafa, A. Guvendi, and H. Hassanabadi,
Optics in spiral dislocation spacetime: Torsion as a geometric waveguide and frequency-filtering mechanism,
Eur. Phys. J. C (2025), \url{https://doi.org/10.1140/epjc/s10052-025-15239-x}, arXiv.2507.14865.

\bibitem{s1}
S. H. V{\"o}lkel, Bound States of the Schwarzschild Black Hole,
Phys. Rev. Lett. {\bf 134}, 241401 (2025),
\url{https://doi.org/10.1103/tbm2-gzv9}.

\bibitem{s2}
S.-M. Wu, H.-Y. Wu, Y.-X. Wang, J. Wang, Gaussian tripartite steering in Schwarzschild black hole,
Phys. Lett. B {\bf 865}, 139493 (2025),
\url{https://doi.org/10.1016/j.physletb.2025.139493}.

\bibitem{s3}
A. Al-Badawi, S. Shaymatov, Y. Sekhmani, Schwarzschild black hole in galaxies surrounded by a dark matter halo,
JCAP {\bf 2025}, 014 (2025),
\url{https://doi.org/10.1088/1475-7516/2025/02/014}.


\bibitem{s4}
A. A. Ara{\'u}jo Filho, S. Zare, P. J. Porf{\'\i}rio, J. K{\v{r}}{\'\i}{\v{z}}, H. Hassanabadi, Thermodynamics and evaporation of a modified schwarzschild black hole in a non--commutative gauge theory,
Phys. Lett. B {\bf 838}, 137744 (2023),
\url{https://doi.org/10.1016/j.physletb.2023.137744}.


\bibitem{s5}
Z. L. Wang, E. Battista, Dynamical features and shadows of quantum Schwarzschild black hole in effective field theories of gravity,
Eur. Phys. J. C {\bf 85}, 304 (2025), \url{https://doi.org/10.1140/epjc/s10052-025-13833-7}.

\bibitem{s6}
F. Javed, Stability and dynamics of scalar field thin-shell for renormalization group improved Schwarzschild black holes,
Eur. Phys. J. C {\bf 83}, 513 (2023), \url{https://doi.org/10.1140/epjc/s10052-023-11686-6}.

\bibitem{s7}
G. Mustafa, F. Atamurotov, I. Hussain, S. Shaymatov, A. {\"O}vg{\"u}n, Shadows and gravitational weak lensing by the Schwarzschild black hole in the string cloud background with quintessential field,
Chin. Phys. C {\bf 46}, 125107 (2022), \url{https://doi.org/10.1088/1674-1137/ac917f}.

\bibitem{s8}
R. C. Pantig, Apparent and emergent dark matter around a Schwarzschild black hole,
Phys. Dark Univ. {\bf 45}, 101550 (2024), \url{https://doi.org/10.1016/j.dark.2024.101550}.

\bibitem{s9}
N. Heidari, H. Hassanabadi, Investigation of the quasinormal modes of a Schwarzschild black hole by a new generalized approach,
Phys. Lett. B {\bf 839}, 137814 (2023), \url{https://doi.org/10.1016/j.physletb.2023.137814}.

\bibitem{s10}
S. Hod, Bound-state resonances of the Schwarzschild black hole: Analytic treatment,
Phys. Rev. D {\bf 112}, 064040 (2025), \url{https://doi.org/10.1103/l5dw-ddrw}.

\bibitem{photon-sphere}
C. M. Claudel, K. S. Virbhadra, G. F. R. Ellis, The geometry of photon surfaces,
JMP {\bf 42}, 818--838 (2001), \url{https://doi.org/10.1063/1.1308507}.

{\color{black}
\bibitem{33} G. Cognola and P. Lecca,
``Electromagnetic fields in Schwarzschild and Reissner--Nordstr\"om geometry. Quantum corrections to the black hole entropy,''
\textit{Phys. Rev. D} \textbf{57}, 1108--1111 (1998), \url{https://doi.org/10.1103/PhysRevD.57.1108},
arXiv:hep-th/9706065.}

{\color{black}
\bibitem{34}
L. C. B. Crispino, A. Higuchi, and G. E. A. Matsas, ``Quantization of the electromagnetic field outside static black holes and its application to low-energy phenomena,'' \textit{Phys. Rev. D} \textbf{63}, 124008 (2001); Erratum: \textit{Phys. Rev. D} \textbf{80}, 029906 (2009), \url{https://doi.org/10.1103/PhysRevD.80.029906}, arXiv:gr-qc/0011070.}

{\color{black}
\bibitem{35}
P. Schneider, J. Ehlers, and E. E. Falco, \textit{Gravitational Lenses}, Springer, 1992.}}
\end{thebibliography}
\end{document}